\documentclass[a4paper,english,12pt,dvips]{article}

\usepackage{graphicx,amsbsy,amsmath,rotating}
\usepackage[english]{babel}
\usepackage{natbib}

\begin{document}

\thispagestyle{empty}

\vskip 15.0cm

\begin{center}

{\Large \LARGE\bf 
A physical model for aftershocks\\
triggered by dislocation \\
\vskip 0.6cm
on a rectangular fault 
}

\end{center}

\vskip 5.0cm

\begin{center} 
\Large \textbf{{\emph{Rodolfo Console and Flaminia Catalli}}}\\
\large\emph{Istituto Nazionale di Geofisica e Vulcanologia, Rome, Italy\\e-mail: console@ingv.it, e-mail: catalli@ingv.it}
\end{center}

\vskip 5.0cm

\begin{center} 
{\Large \textit{June 2005\\
Submitted to Special Issue Tectonophysics\\
EGU-NP3.06 Session Vienna}}
\end{center}

\section*{Abstract}

We find the static displacement, stress, strain and the modified Coulomb failure stress
produced in an elastic medium by a finite size rectangular fault after its
dislocation with uniform stress drop but a non uniform dislocation on the source. The time-dependent rate of triggered
earthquakes is estimated by a rate-state model applied to
a uniformly distributed population of faults whose equilibrium is
perturbated by a stress change caused only by the first dislocation. The rate of triggered events in our simulations is exponentially
proportional to the Coulomb stress change, but the time at which the maximum rate begins to
decrease is variable from fractions of hour for positive stress changes of
the order of some $MPa$, up to more than a year for smaller stress changes. As
a consequence, the final number of triggered events is proportional to the Coulomb
stress change. The model predicts
that the total number of events triggered on a plane containing the fault is proportional to the $2/3$ power of the seismic moment. Indeed,
the total number of aftershocks produced on the fault plane
scales in magnitude as $10^{M}$. Including the negative contribution of the
stress drop inside the source, we observe that the number of
events inhibited on the fault is, at long term, nearly identical to the
number of those induced outside, representing a sort of conservative natural
rule. Considering its behaviour in time, our model doesn't completely  match the popular 
Omori law; in fact it has been shown that the seismicity induced closely
to the fault edges is intense but of short duration, while that expected at
large distances (up to some tens times the fault dimensions) exhibits a much
slower decay.\\
\\
\textbf{Key Words:} aftershocks, earthquake triggering, fault parameters, rate-and-state, Omori law.  

\section{Introduction}

Studying the space-time interaction of earthquakes is important not only for
the comprehension of the phenomenon of earthquakes, but also for its
possible application in the mitigation of earthquake risk.

In this paper we aim to the physical modelling of the interaction between
seismic events, seeking a relationship between the source parameters of an
earthquake and the rate of all other earthquakes that follow it, a
phenomenon empirically and variably labelled with the names of seismic
acceleration, foreshock and aftershock sequences, swarms, earthquake
clustering, etc.

A purely statistical approach to this phenomenon exists with the name of
''epidemic model'' \citep{ogata88, ogata98, console03}.
In the epidemic model each earthquake source is supposed capable of
increasing the probability of new earthquakes according to a kernel space
and time probability distribution decreasing from the source. The epidemic model is applicable not only to the
aftershock phenomenon but also to the foreshock's. However, despite it provides an accurate statistical rapresentation of earthquake interaction, the physical interpretation is ignored.

In order to formulate a more physical description of the clustered seismicity,
we consider that seismic events modify the stress field around the causative fault. 
Our main goal in this work is to observe the spatio-temporal variation of the seismicity rate after an event introducing in the algorithm a rate-state model approach derived by  \citet{ruina83}
and Dieterich \citep{dieterich86,dieterich92,dieterich94}. In particular, we seek a physically based relationship between the magnitude of the causative fault and the number of its aftershocks. We aim to create a link between a physical modelling of the stress transfer and the epidemic model approach (see also \citet{console04}).
Some recent studies put in evidence that sudden stress variations, even of
small magnitude, can produce large variations of the seismicity rate. It is
recognized as a phenomenon of triggering, rather than induction, on faults
that are already subject to relevant stress. The seismic rate increases in
general where the stress change (called \emph{Coulomb Stress Change)} is positive,
according to the Coulomb model \citep{mendoza88, boatwright96, stein97, gomberg98, gomberg00, harris98, king01, stein99, toda00, kilb02, belardinelli03, nostro05}. These studies were able to give a physical
interpretation for earthquake interaction observed in specific real cases,
but were not suitable for the general application of the model in predictive
way as it is necessary for the information contained in a seismic catalog.
In particular, the Coulomb stress change criterion doesn't allow to model
the hyperbolic time decay of the aftershock rate after a mainshock, known as
the \emph{Omori Law}. In this simple model all the next earthquakes would be advanced by a time wich depends on the stress change and on the loading rate but doesn't depend on the physical properties of the fault. The
experience shows that the Earth doesn't rather behave in this way. The most
popular empirical description of the phenomenon is the Omori law, stating
that the the aftershock rate decays in time as $t^{-1}$. Among the various
theories modelling the Omori law, we have taken in consideration only the
one that involves a particular constitutive behaviour of faults.
Therefore, we have not considered, even without rejecting them in principle,
other hypotheses that predict a time variation in stress, like the
viscoelasticity and the diffusion of fluids in the medium. The
rate-state model for earthquake nucleation seems capable of substantially
explaining all the phenomenology. In this model slip doesn't occur
instantaneously at the exceeding of a threshold stress as for the
Coulomb-Amonton criterion, but follows a more complex time history with
different phases. This time history depends on the physical properties of the fault too.
\citet{gomberg05a} have shown that the rate-state Dieterich model may be derived from simple general ideas. 
The application of the rate-and-state model to the evolution of seismicity is not new, e.g., \citet{toda98} have used this model to predict the spatio-temporal seismicity rate after the Kobe earthquake. Other examples of application of the same model can also be found in the literature \citep{toda00, toda02, toda03}. Following the rate-state model it has recently been
possible to simulate seismicity quite realistically, accounting for the rate
of events induced by stress changes even at large distance from the inducing
earthquake \citep{ziv03,ziv-rubin03}.

Basing our work on the classical theory of the elasticity, we start from the
formulation of the stress released by a rectangular fault with uniform
stress-drop; then we apply the rate-state model \citep{dieterich94} obtaining the complete time and space
distribution of the seismicity induced by a given fault. It is important to remark that for present purposes we neglect the interaction between subsequent events, supposing that only the first event perturbs the stress field. We also neglect the effect of the free surface of the Earth. In this study we limit our interest to the behaviour of induced seismicity in space and time and its relationship with the fault's parameters. Because of the qualitative nature of this work we don't use the expressions derived, in a similar but more complex context, by \citet{okada92}. The results of
simulations are analysed to find out the scaling relationships existing
between the free parameters of the model and the expected seismicity in a way that will allow the validation of the model by real observations.

\section{Elastic model for a rectangular fault}

The theoretical approach of this work has been made as simple as possible, still preserving the capability of modelling the stress transfer from a seismic source to another.

Suppose that a fault is embedded in a homogeneous, isotropic, elastic medium
where the stress tensor $\bf{\sigma}$ is uniform. At a particular
moment, the fault slips generating an earthquake, and suppose that the
earthquake fully releases the component of the traction parallel to the slip
vector across the fault. We call $\Delta \sigma$ (the stress drop) the uniform
negative change of traction on the fault. It is well known that the slip
distribution on a fault in similar conditions is not uniform. The slip
distribution will satisfy the theory of the elasticity at the new
equilibrium, and it will be zero on the edges of the fault. It is also well
known that the shear stress on the fault plane outside the edges of the
fault increases significantly.

The analytical solution for the slip distribution on a fault at the
equilibrium doesn't exist except that for very simple geometries, such as a
rectangular fault of constant width and infinite length \citep{knopoff57} and for a circular fault \citep{borok59, udias99}. As mentioned by \citet{kostrov98}, the analytical
solution is not known even for a geometry as simple as a rectangular shape
of width $W$ and length $L$. In the literature it is possible to find the
computation of the stress field generated by an ideal rectangular fault with
uniform slip in an infinite homogeneous space or semispace \citep{chinnery61, iwasaki79, okada85, okada92}.
However, as mentioned above, this is not a situation of equilibrium
fulfilling all the physical constraints imposed by the theory of the
elasticity. For this reason, here we make use of a non uniform slip
distribution that is approximatively compatible with a uniform stress drop on the fault. 

To calculate the static displacement, stress and strain, we start from the components $i$ of the static displacement vector $u_{i}^{j}
\mathbf{(x)}$ produced by an elementary force directed as the $j$ axis in
any point of an infinite, elastic, homogeneous and isotropic medium, known
as the Somigliana Tensor. From the expressions valid for a single force
source, those for a couple of forces of given moment are derived by
differentiation with respect to the direction of the arm of the couple of forces. Then the
expressions for a source represented by a double couple of forces are
obtained by summing a pair of two co-planar couples orthogonal to each
other. Once the expressions for the displacement are known, the stress $\sigma _{ij}$, the strain $e_{ij}$ and the tractions are computed by their definitions. Indeed we also can compute the modified Coulomb failure stress from the relation:
\begin{equation}
\sigma_{c}=T_{s}-\mu'T_{n},
\end{equation}
where $T_{s}$ and $T_{n}$ are respectively the traction parallel to the slip vector and the traction normal to the fault plane (negative for compression); 
$\mu'$ is the effective coefficient of friction incorporating the pore fluid pression that is commonly assumed to be proportional to normal stress change. The effective coefficient of friction in coseismic stress change studies has varied from $0.0$ to $0.75$, with an average value of $\mu'=0.4$ often used. This is the value we have used in our simulations.

Our model of an extended source is drawn euristically from that of a rectangular fault infinitely
extending in one direction and that of a circular fault, and it is
characterized by a slip distribution as:
\begin{equation}
\Delta u=\frac{\Delta \sigma }{\mu }\sqrt{\frac{\left[ \left( \frac{W}{2}
\right) ^{2}-x^{2}\right] \left[ \left( \frac{L}{2}\right) ^{2}-y^{2}\right]
}{\frac{LW}{4}}},  \label{eq:deltau}
\end{equation}
defined for $-{W}/{2}\leq x\leq W/2$ e $-{L}/{2}\leq y\leq L/2$, where $x,y$
are the coordinates on the fault plane whose origin is coincident with the
center of the rectangle, $\Delta \sigma $ is the stress drop and $\mu $ the
shear modulus (rigidity) of the elastic medium. Equation (\ref{eq:deltau})
satisfies the condition of zero slip on the edges of the rectangle. The slip
distribution is of a form very similar to that obtained numerically by
discretization, except for the cases when one dimension is much larger than
the other \citep{kostrov98} . Its analytical form is
similar to the solution for a circular fault. Moreover, for a fault of
square shape ($W=L$) it achieves a seismic moment consistent with those
obtained analitically for two circular faults, respectively inscribed and
circumscribed to the square. \citet{bonafede00} have shown that, when imposing a unidirectional traction release in the 
strike or dip direction, a minor component of slip is present, over the fault plane, even in the 
direction perpendicular to the released traction. In consideration of the qualitative character of
this study, we have neglected this component.

For our numerical applications we discretize the continuous slip
distribution of equation (\ref{eq:deltau}) with a set of point sources
densely distributed on the fault. The more numerous are the point sources,
the more accurate will be the approximation. Each element $(p,q)$ of the
fault is characterized by its scalar seismic moment $m_{pq}$ that, by the
theorem of the representation is given by:
\begin{equation}
m_{pq}=\mu \Delta u_{pq}\Sigma _{pq},
\end{equation}
where $\Delta u_{pq}$ is the slip on the element $(p,q)$ and $\Sigma _{pq}$
is its elementary area.

If $\mathbf{u}_{pq}(\mathbf{x})$ is the displacement produced by the slip on
the element $(p,q)$ in the point $\mathbf{x}$, the displacement resulting
from the combination of all the elements is computed by:
\begin{equation}
\mathbf{u}(\mathbf{x})=\sum_{p=1}^{M}\sum_{q=1}^{N}\mathbf{u}_{pq}(\mathbf{x}
),
\end{equation}
where $\mathbf{u}_{pq}(\mathbf{x})$ is computed by the relation valid for a
point source of moment $m_{pq}$.

Following equation (\ref{eq:deltau}) the slip $\Delta u_{pq}$ is written as:
\begin{equation}
\Delta u_{pq}=\Delta u_{max}\frac{\sqrt{\left[ \left( \frac{L}{2}\right)
^{2}-x_{1p}^{2}\right] \left[ \left( \frac{W}{2}\right) ^{2}-x_{3q}^{2}
\right] }}{\frac{LW}{4}},  \label{eq:deltaudiscreto}
\end{equation}
where, in this case, $x_{1p}$ and $x_{3q}$ are the $x_{1}$ and $x_{3}$
coordinates of the element $(p,q)$ respectively, and:
\begin{equation}
\Delta u_{max}=\frac{\Delta \sigma }{\mu }\frac{\sqrt{LW}}{2}
\label{eq:ricorda}
\end{equation}
is computed through the other three fault parameters. These parameters are
also connected to the scalar moment $M_{0}$ through the relations

\begin{equation}
M_{0}=\mu \overline{\Delta u}WL=\frac{\pi ^{2}}{16}\mu \Delta u_{max}WL=
\frac{\pi ^{2}}{32}\Delta \sigma (WL)^{3/2}  \label{Relaz_M0_LW}
\end{equation}
where $\overline{\Delta u}$ is the average slip on the fault.

The above mentioned theoretical approach has been implemented in a computer
code which allows the user to chose among several options:
- The focal mechanism characterizing the causative fault and the induced events 
constrained by the strike $\phi$, the dip $\delta$ and the rake $\lambda$ angles;
- Dimensions and stress drop, seismic moment, or magnitude of the fault;
- Type of output: Displacement (3 components), stress or strain (6
independent components), traction parallel to the slip vector and perpendicular to the fault plane (2 component) and Coulomb failure stress (1 component).

The output values are given on a rectangular grid of points, whose
orientation (horizontal, or vertical parallel to one of the co-ordinate
axes), dimensions, and spacing are also selected by the user.
See Figure 1 for the conventions we adopted for the reference system and the angles. 

In this work we make, for sake of simplicity, the assumption that all the aftershocks have the same focal mechanism of the causative fault.

Figures 2 and 3 show the component
of the stress field parallel to the slip $\sigma_{12}$ and the normal traction $T_{n}$ produced by a vertical 
strike slip fault whose parameters are
listed in Table \ref{tab:simulazioni1}. Note that these figures show the
stress change and not its absolute value.

Figure 2 clearly shows that the shear stress $\sigma _{12}$ is
maximum near the edges of the fault, in particular along the slip direction $
x_{1}$, and decreases with increasing the distance from the fault. Inside
the fault the stress change is negative and approximately equal to $\Delta
\sigma$. 

Figures 4, 5 and 6 show respectively $\sigma_{12}$, $T_{n}$ and the Coulomb failure stress $\sigma_{c}$
produced by a vertical strike slip fault on the horizontal plane.

\section{The rate-state model}

As already stated, we assume that the time-dependent behaviour of the
seismicity triggered by a Coulomb stress change in a population of faults
is described by the rate-state model introduced by \citet{ruina83} and Dieterich \citep{dieterich86, dieterich92, dieterich94}. According to \citet{dieterich94}
the rate $R$ of earthquakes due to a $\Delta \tau $ Coulomb failure stress is given
by:
\begin{eqnarray}
R &=&\frac{r\frac{\dot{\tau}}{\dot{\tau _{r}}}}{\left[ \frac{\dot{\tau}}{
\dot{\tau _{r}}}\exp \left( \frac{-\Delta \tau }{A\sigma }\right) -1\right]
\exp \left( \frac{-t}{t_{a}}\right) +1}\qquad \dot{\tau}\neq 0,  \nonumber
\label{eqnarray:Dieterich} \\
&&
\end{eqnarray}
whose various parameters are defined in Table \ref{tab:variabili}.
Note that equation (\ref{eqnarray:Dieterich}) doesn't predict the magnitude
of the earthquakes, for the distribution of which we assume the classical
\emph{Gutenberg-Richter }frequency-magnitude relationship. However, both $R$
and $r$ refer to the number of earthquakes in a specific magnitude range.

The modelling of the space-time distribution of seismicity triggered by
an earthquake is straightforward, introducing the expression for the Coulomb failure stress
$\sigma_{c}$ in equation (\ref{eqnarray:Dieterich}) for $\Delta\tau$. We make the assumption $\dot{\tau}
=\dot{\tau _{r}}$. The meaning of this assumption is that the primary
earthquake doesn't change the long-term stress rate due to the tectonic
driving forces in the environment.

It is easily understandable from (\ref{eqnarray:Dieterich}) that the
integral of $R$ over infinite time diverges. This is due to the fact that
the limit of $R$ for $t\rightarrow \infty $ is the background rate $r$. We
prefer to deal with the net triggered seismicity rate $R^{\prime }=R-r$. The
plot of $R^{\prime }$ versus time for different values of $\Delta \tau $ is
shown in Figure 7 (see also \citet{dieterich94}).
It shows that the initial value of the triggered seismicity rate, immediately after the triggering event, is proportional to the exponential of the stress variation but the time period over which $R(t)$ is maximum and relatively constant decreases exponentially by the same factor for increasing Coulomb stress change. This time period, in substantial agreement with \citet{dieterich94}, can be obtained computing the intersection between $R^{\prime}(t=0)$ and the straight line corresponding to the limit of $R^{\prime}$ when $\Delta\tau\rightarrow\infty$. The difference with \citet{dieterich94}, as Figure 8 shows, is that we have preferred to make use of $R^{\prime}$, rather than $R$, and we ignore the case $\dot{\tau}=0$ which we judge not realistic. Such time period ranges from few hours for a variation $\Delta\tau =5MPa$ to nearly one year for $\Delta\tau=1MPa$. Figure 7 finally shows that the total duration of the triggered seismicity is the same but the shape of the decay is different for different $\Delta\tau$. 

The integration of $R'$ over infinite time leads to the not completely trivial result thet the total number of triggered events in an area of constant stress change is proportional to $\Delta\tau$ \citep{helmstetter04}:

\begin{equation}
\int_{0}^{\infty }R^{\prime }(t)dt=\frac{rt_{a}}{A\sigma }\Delta \tau =\frac{
r}{\dot{\tau}}\Delta \tau.  \label{integr_R}
\end{equation}

This relationship predicted by the friction model had been observed for Landers Aftershocks by \citet{gross97}.

Consider the integral over time of the
triggered seismicity rate $R^{\prime}$, once for a positive $\Delta \tau $, and then for
a negative $\Delta \tau $ of the same size, and we sum the two functions. We obtain a quantity that intially has a positive
increase and then tends back to zero (Figure 9). This
result can be interpreted as a sort of conservative law of the seismicity,
by saying that the total number of aftershocks generated in a place close to
a fault can be balanced, at long range, by the total number of those events
that are inhibited by the stress drop in a place internal to the fault if the stress drops are equal in
absolute value. It seems to us that this circumstance has been ignored, so far, in all previous studies on this subject. 

For numerical applications it is necessary to define the value of the
various parameters appearing in equation (\ref{eqnarray:Dieterich}). Table \ref{tab:parametri} contains
a list of values inferred from several
geophysical considerations and used in our simulations. The reference
seismicity rate, $r$, is a quantity that should be known experimentally for
a given area, and it is related to the average strain rate $\dot{\tau}$ of
that area. We use for $r$ a uniform value of four events of magnitude exceeding
3.0 per year per 1000 $km^{2}$, and for $\dot{\tau}$ a value of 5 $KPa$ per
year. These are simplifications of the reality where $r$ and $\dot{\tau}$ are not geographically uniform.
These parameters seem however reasonable for an area of moderate seismicity.
The reason why $r$ is expressed as a number of events per unit area and time
is given below. The parameter $A$ of the constitutive law has a value
ranging from 0.005 to 0.015 obtained from laboratory experiments \citep{dieterich95}.
Generally, simulations that most nearly resemble earthquakes in nature were obtained with rather small
values of $A$, near to $A\approx0.001$. The value of $A\sigma$ has been evaluated for different sequences
of earthquakes by several autors \citep{belardinelli03, stein99}.
The parameter $\sigma $ is the normal pressure on the
fault (of the order of some tens of $MPa$). The characteristic time of aftershocks, $t_{a}$, depending
on the other parameters through the
relation $t_{a}=(A\sigma )/\dot{\tau}$, in this context is of the order of
several tens of years. This value is essential to model a long term
triggered seismicity, because the effect of the stress change disappears
completely after a time of the order of the double of $t_{a}$.

\section{Application to a rectangular fault}

\citet{dieterich94} discussed in detail the case of a finite circular dislocation with uniform stress drop.
In our simulations we have taken into consideration a vertical rectangular
fault with horizontal slip in the $x_{1}$ direction, embedded in a
homogeneous, isotropic, elastic medium. The values assumed for the fault
parameters are again those reported in Table \ref{tab:simulazioni1}.
Assuming a constant stress drop $\Delta \sigma$, equation (\ref{Relaz_M0_LW}) shows that the scalar seismic moment $
M_{0}$ is proportional to the quantity $(WL)^{3/2}$, that can be considered
a sort of equivalent source volume. Scaling $W$ and $L$ by a given factor, $
M_{0}$ scales accordingly, and also the shape of the stress distribution in
space scales by the same factor. Assuming that the background seismicity
rate is given as the number of events per unit volume and time, and
integrating over the space, we find that the number of events triggered by
the earthquake is proportional to the cube of the linear dimensions, i.e. to
the seismic moment $M_{0}$, or to $10^{\frac{3}{2}M}$. In this study we make the
different hypothesis that the aftershocks of an earthquake occur mainly on
the fault plane. This hypothesis comes from the consideration that the fault
plane is subject to the maximum sheare stress and that it is a sort of plane
of weakness where the triggered seismicity is more likely to occur.
In this case, the number of triggered events would be proportional to $M_{0}^{2/3}$,
or to $10^{M}$. If it was true, and neglecting other kinds of factors, one
earthquake of given magnitude would have, all togheteher, a number of aftershocks ten times
larger than one earthquake of one magnitude unit smaller. Assuming, then,
that the $b$ value of the Gutenberg-Richter frequency-magnitude relation is
equal to 1, it would follow that the earthquakes of a magnitude class
produce the same number of aftershocks as the earthquakes of the magnitude
class one unit smaller. This hypothesis is supported by the statistical
analysis of real observations \citep{Felzer02, yamanaka90, utsu69} and will be used in the following of the paper.

A first group of simulations concerning the seismicity triggered by a fault
as a whole have been performed to study the variation of the total number of
triggered events (in space and time), $N_{tot}$, versus the main free source
parameters: the stress drop $\Delta \sigma $, the linear dimensions $L$ e $
W $ and the magnitude $M$. We assume that the magnitude $M$ is proportional to the logarithm of the seismic
moment $M_{0}$ (measured in $Nm$) according to the \citet{kanamori75} relation:
\begin{equation}
\log M_{0}=9.1+1.5M .
\end{equation}

Tables \ref{tab:sigmafisso}, \ref{tab:LWfisso}
and \ref{tab:Mfisso} show the results obtained assigning a fixed value
for one of the parameters at a time, and letting the others change. These
simulations have been carried out by summation of the single contributions
of the elementary cells constituting the area surrounding the fault. Only
the seismicity triggered on the fault plane, but outside the fault edges,
has been considered.
Looking at Table \ref{tab:sigmafisso} one can notice that, under our
hypoteses, for a constant stress drop, the total number of triggered events
increases about proportionally to the fault area $LW$. Theoretically it should be exactly proportional to the area of the source, but in the calculations there are numerical imprecisions due to the grid size approximations. These numerical simulations
show, also, that the number of aftershocks is proportional to the 2/3 power
of the scalar seismic moment or to the exponential of the magnitude ($10^{M}$),
 in agreement with the above theoretical arguments. Table \ref
{tab:LWfisso} clearly shows the proportionality existing between the stress
drop $\Delta \sigma $ and $N_{tot}$, having fixed the linear dimensions of
the fault. Furthermore, the results reported in Table \ref{tab:Mfisso}
give credit to the hypothesis that, for a constant magnitude, the
earthquakes of smaller linear dimensions generate a larger number of
triggered events.

Let's mention, lastly, the results of a simple numerical test. We just have
theoretically shown in the previous section that the total number of events triggered by a stress
change $\Delta \tau $ per unit area is, after a time much larger than the
characteristic time $t_{a}$, identical to the number of events inhibited by
a negative Coulomb stress change $-\Delta \tau $ on the same area. The numerical
simulations about an extended source reflect this idea in a more general way. In fact, if we take the
total number of aftershocks produced in a time interval of 100 years by a
fault of $100$ $km^{2}$ on the portion of an area of $10^{4}$ $km^{2}$
external to it, we find $353.2$ events, while the same computation made
including the negative contribution of the area internal to fault itself
gives only $15.6$ (less than $5\%)$. The idea of an overall null balance is, then,
substantially met: the phenomenon of clustering doesn't change the long term
seismicity rate of a large area. This is in contrast with the opinion that trigger zones produce many more shocks than are missing from the shadows \citep{toda03}. We neverthless regard for \citet{marsan03} that put in evidence the practical difficulty to observe in nature the lack of events (the quiescence) mainly when looking at weakly active regions and short timescales. 

\section{Temporal behaviour of the triggered seismicity}

Let's now consider in more detail the dipendence on time of the number of
triggered events predicted by the present hypothesis. \citet{kagan91} and \citet{dieterich94} found, in their empirical works, that shallow aftershocks, above a given magnitude cutoff, decay globally (for different distance intervals) in about $10$ years and this time decay appears to sistematically decrease with increasing depth.
In our model the behaviour in
different zones depends on the different values of the stress changes, that
decrease with the increasing distance from the edges of the fault. It can be
then noted that the plateau of the plot of the seismicity rate in time (see Figure 7) has a duration that increases with the
increasing distance from the source. This result depends on our assumption of considering only
aftershocks on the mainshock fault plane. It can also be biased by the fact that we consider only
the effect of static stress step $\Delta \tau $ and we ignore the interaction between events. In spite of that, we believe that our model
can decribe a behavior not so far from reality. Therefore we keep  as our reference the model of
fault given in Table \ref{tab:simulazioni1} again.

As shown in Figure 10a, in the close neighbouring of the fault
(within $0.5$ $km$ from the perimeter of the fault), most of the
aftershocks occur within one day after the occurrence of the inducing
earthquake, and more than $1/3$ of them occur already in the first hour.
This is clearly not the case for further distance from the source: in the
slice between $0.5$ and $8.5$ $km$ from the fault, about half of the
activity is exploited within about one year. The further we go apart from
the source, the longer is the time interval necessary for the exploitment of
the triggered activity. In fact, in the distant zone (from $20$ to $100$ $km$
from the fault) most of the triggered events occur longer than 20 years
after the inducing earthquake (Figure 10b) .  
The behaviour shown in Figure 10a and 10b is illustrated also by the
snapshots of the spatial distribution of the seismicity rate at different
times. Figures 12, 13 and 14 clearly show how the distribution of aftershocks resembles
that of the Coulomb stress change (Figure 2) and how the time decay is different in the
different slices. These also show how shadow zones (where the seismicity is inhibited) have an effect longer in time.
It should be noted again that we make the assumption that the aftershocks occur mainly on the fault plane and 
that these have the same focal mechanism of the causative event.

We are now interested to check if the generalized Omori law is suitable to
describe the temporal behaviour of $R^{\prime }(t)$ predicted by our model
in every spatial slice and, in such case, by means of which free parameters.
We refer to the generalized Omori law as to the formula describing the decay
of aftershock rate after a mainshock \citep{utsu95}
through three free parameters $a$, $c$ e $p$:
\begin{equation}
R^{\prime }(t)=\frac{a}{(c+t)^{p}}.  \label{eq:OmGen}
\end{equation}
For this purpose we make use of an algorithm for the least squares best-fit
of the sets of data reported in Figure 10 by the integral of (\ref{eq:OmGen}) over the time:
\begin{equation}\label{eq:OmInt}
y=\int_{0}^{t}R^{\prime }(t^{\prime })dt^{\prime }=a\left[ \frac{\left(
c+t\right) ^{1-p}-c^{1-p}}{1-p}\right] .
\end{equation}

The results obtained for the best-fit are shown in Table \ref{tab:rgromori}. 
It is evident that the values of the parameters obtained are quite
different from case to case (for instance, the $c$ parameter is extremely
small for the slices closest to the fault edges), and that the standard
deviation is larger for the slices that include the wider range of distances
from the fault. These results show that the same set of parameters in
equation (\ref{eq:OmInt}) can not fit the synthetic values obtained for
different zones. The observation that the Omori $p$ parameter varies spatially in real aftershocks sequences has been documented in a number of papers \citep{wiemer99, enescu02, wiemer00, wiemer02, wiemer-wyss02}.
\citet{wiemer02} reported a case of significantly different $c$ values in the northern and southern Hector Mine aftershock volume, respectively. \citet{enescu02} in their study of aftershock activity of the 2000 Western Tottory earthquake found c-values very close to $0$ except for larger magnitude earthquakes, arguing that this feature could result from incompleteness of data, or might also reflect the complexity of the rupture process.

The situation is illustrated also by the plots of Figures 15 and 16, showing the graphical comparison between the synthetic data and the relative
theoretical approximation for different slices.

\section{Conclusions}

In the development of this model some very crude approximations have been made.
Two of the most important are (a) having neglected the dishomogeneneity of the
stress drop on the fault, and consequently the very complicated pattern of
slip (this must certainly have influence on the aftershock pattern: some
aftershocks may even occur inside the fault, where the gradient of slip is
high); (b) having neglected the interaction between subsequent events.  

The first step of our modelling, concerning the stress change in the
medium following an earthquake, has given results comparable to those of
previous papers as, for instance, \citet{chinnery61}, \citet{king01}, \citet{belardinelli03}. This substantial agreement,
though the simplifications introduced in this specific case of rectangular
source, supports the validity of our methodology.

As to the interaction among seismic events, the application of the rate-state model of
\citet{dieterich94} has shown how the total number of triggered events per unit area
is proportional to the Coulomb stress change. Nevertheless, the time constant by
which the rate of events decreases has a large variability, ranging from a
fraction of hour for a Coulomb stress change of several $MPa$, up to tens of years
for stress variations smaller than 1 $MPa$. 

A first set of simulations
carried out with a model of rectangular fault, based on the hypothesis that
most of the aftershocks occur on the same plane of the fault, has shed light
on the relationship between the main source parameters and the number of
triggered events. The simulations lead to the conclusion that many
earthquakes of small magnitude can produce, in total, a number of
aftershocks comparable to that of fewer larger earthquakes. Our model
predicts, in fact, that, for a constant stress drop, the number of triggered
events is roughly proportional to the seismic moment at the power of $2/3$,
or equivalently to $10^{M}$.

An important aspect of the model concerns
the time-space behaviour of the triggered seismicity. The simulations show
that the application of the rate-state model and equation (\ref
{eqnarray:Dieterich}) introduced by \citet{dieterich94} to the static
model of stress change has as a consequence something more complicate than
the conventional Omori law for the temporal decay of aftershock rate. In
fact, the seismicity is expected to be intense (in accordance to the stress
change,) and have a maximum constant value, close to the fault edges of the primary earthquake, but the time necessary for the starting of decay is shorter. On the contrary, the seismicity expected at
larger distances (up to 10 times the linear dimensions of the fault) is
weak, but the time necessary for the decay from the maximum value is longer (even some tens of years). We can conclude that the temporal behaviour of aftershock rate varies in space and this is theoretically because "c" increases with the distance from the fault edge. Moreover, the
model predicts that the total long term production of aftershocks, including
wider areas, is not negligible respect to the short term production. In this
respect, the regional background seismicity could be interpreted as a sort of noise,
or memory, due to the superposition of the effect of many older earthquakes.
The time decay described by the popular Omori law could be interpreted as an
apparent average result of the contribution from the various areas on the
plane containing the primary fault.

Let's list, in short, the most important results concerning the static
effect of a dislocation, as following:

\begin{itemize}
\item  The number of triggered events is proportional to the area of the
fault, for a constant stress drop;

\item  the number of aftershocks is proportional to the $2/3$ power of the static seismic
moment, for a constant stress drop;

\item  the logarithm of the total number of aftershocks scales linearly with
the magnitude, for a constant stress drop;

\item  the number of triggered events is proportional to the stress drop on
the fault, if its linear dimensions are kept constant.
\end{itemize}

The rate-state model \citep{dieterich94} used in our algorithm
for the time distribution of the seismicity produced by a dislocation on
a rectangular fault, has allowed a partial justification of the Omori law.
We outline the following main conclusions obtained from the simulations:

\begin{itemize}
\item  The seismic activity is very intense during the first few hours or
days after the occurrence time of the primary earthquake, in the rectangular
slice close to the edges of the fault;

\item  The immediate decay of this intense activity is followed by a period
of time during which the activity at larger distance (of the order of the
linear dimensions of the fault) is nearly constant. For longer time (several
years) the total number of triggered events in the external zone becomes
comparable with that of the most internal one.

\item  the number of events triggered over a long time scale at distances
larger than the fault linear dimensions is significantly large;

\item  the seismic activity inside the fault drops to negligible values at the occurrence of the primary
earthquake and returns to normal values only after a time much longer than
the characteristic time (several tens of years in our simulations); for a
long time range the number of events inhibited inside the fault is
comparable to that of the events triggered outside: the net balance can be
considered null; the seismicity rate changes in space and time, but it
doesn't have influence on the rate averaged over very large intervals of space and time.
\end{itemize}

In light of the consequences derived from our model, both positive and negative perturbations of the seismic activities resulting from the interactions among earthquakes would last for a time of the order of decades. In this respect, it seems unrealistic to define  and observe  what is commonly called "`background seismicity"' of an area.
 
We can conclude that the model here analysed, though its very simple
conception, gives a physical justification both to the very popular
phenomenon of short term earthquake clustering (aftershocks and foreshocks
in strict sense) and to that of the long term induction, also observed in
various occasions. 

We are confident that the refinement of the model and its validations with real catalogs will bring to even more interesting consequences.

\section*{Acknowledgments}
We are grateful to Massimo Cocco for the stimulating discussions and his criticism wich led to a significantly better version of the paper. 
We also would thank Tom Parsons for many useful suggestions to help make this paper clear.

\newpage

\section*{Authors}
 
\indent 
 
\textbf{R. Console},\\ 
Istituto Nazionale di Geofisica e Vulcanologia,\\
via di Vigna Murata, 605\\
00142, Rome, Italy\\
mailto: console@ingv.it\\
phone:+39 06 51860417\\
fax:+39 06 5041181

\vskip 1.0cm

\textbf{F. Catalli},\\
Istituto Nazionale di Geofisica e Vulcanologia,\\
via di Vigna Murata, 605\\
00142, Rome, Italy\\
mailto: catalli@ingv.it\\
phone:+39 06 51860571

\newpage

\section*{Tables}

\begin{table*}[!h]
\caption{Source parameters used in numerical applications.}
\begin{center}
		\begin{tabular}{ll}
		\hline
		Parameter  &
		 \multicolumn{1}{l}{Value}\\
		\hline
	Strike, dip and rake & $0°,90°,0°$\\ 	
	Dimension of the fault & $10$ $km$$\times$$6$ $km$ \\
  Spacing on the fault & $0.1$ $km$ \\
  Dimension of the grid & $30$ $km$$\times$$30$ $km$ \\
  Spacing on the grid & $0.5$ $km$ \\
  Magnitude & $6$ \\
  \hline
	  \end{tabular}
	\end{center}

	\label{tab:simulazioni1}
\end{table*}
\begin{table*}[htb]
\caption{Constitutive parameters used in \emph{Dieterich}'s relation.}
	\begin{center}
		\begin{tabular}{ll}
		\hline
		Symbol  &
		 \multicolumn{1}{l}{Description}\\
		\hline
	$r$ & reference seismicity rate \\
  $\dot{\tau}$ & shear stressing rate \\
  $\dot{\tau}_{r}$ & reference stressing rate \\
  $\Delta\tau$ & earthquake Coulomb stress change \\
  $A$ & fault constitutive parameter\\
  $\sigma$ & normal stress\\
  $t$ & time\\
  $t_{a}$ & characteristic time for seismicity to return to the steady state\\

  \hline
	  \end{tabular}
	\end{center}
   \label{tab:variabili}
\end{table*}
\begin{table*}[htb]
\caption{Values of the parameters used in the numerical applications of the model for induced seismicity.}
	\begin{center}
		\begin{tabular}{ll}
		\hline
		Parameter  &
		 \multicolumn{1}{l}{Value}\\
		\hline
	$r$ & $4$ $events/(y\cdot 1000km^{2})$ \\
  $\dot{\tau}$ & $5$ $KPa$ \\
  $\dot{\tau}_{r}$ & $5$ $KPa$ \\
  $A$ & $0.008$\\
  $\sigma$ & $30$ $MPa$\\
  $t_{a}$ & $48$ $y$\\
  \hline
	  \end{tabular}
	\end{center}

	\label{tab:parametri}
\end{table*}
\begin{table*}[!ht]
	\caption{Relation between fault parameters and aftershock production. The value of $\Delta\sigma$ is
fixed at $4$ $MPa$; the grid used in this computation has an
area of $10^{4}$ $km^{2}$, the spacing on it is $0.5$ $km$. The spacing on the fault is $0.05$ $km$.}
	\begin{center}
		\begin{tabular}{|c|c|c|c|}
		\hline
	$L\cdot W$ $\left[km^{2}\right]$ & $M$ &$M_{0}$ $\left[Nm\right]$  &$N_{tot}$   \\
	\hline
  4 & 4.6 & $9.9\cdot 10^{15}$ &13.4 \\
  25 & 5.4& $1.5\cdot 10^{17} $ & 78.2 \\
  100 & 6.0& $1.2\cdot 10^{18}$  & 263.2 \\
  225 & 6.4& $4.2\cdot 10^{18}$  & 532.8 \\
  \hline
	  \end{tabular}
	  
	\end{center}

	\label{tab:sigmafisso}
	\caption{Relation between fault parameters and aftershock production. The area of the fault is fixed
at the value of $100$ $km^{2}$, the spacing is of $0.05$ $km$; the grid used in the computation has
an area of $17\times 17$ $km^{2}$, the spacing on it is of $0.5$ $km$.}	
\begin{center}
		\begin{tabular}{|c|c|c|c|}
		\hline
	$\Delta\sigma$ $\left[MPa\right]$  & $M$ & $M_{0}$ $\left[Nm\right]$ & $N_{tot}$   \\
	\hline
  1 & 5.6 & $3.1\cdot 10^{17}$ & 51.8 \\
  2 & 5.4 & $6.2\cdot 10^{17}$ & 103.4 \\
  3 & 5.9 & $9.3\cdot 10^{17}$ & 155.2 \\
  4 & 6.0 & $1.2\cdot 10^{18}$ & 206.8\\
  5 & 6.1 & $1.6\cdot 10^{18}$ & 258.6 \\
  \hline
	  \end{tabular}
	\end{center}

	\label{tab:LWfisso}	
\end{table*}

\begin{table*}[!h]
	\caption{Relation between fault parameters and aftershock production. The value of $M$ is fixed at $6.0$;
the output grid is of $20$ $km$$\times$$20$ $km$, the spacing is $0.5$ $km$. The spacing on the fault is $0.05$ $km$.}
	\begin{center}
		\begin{tabular}{|c|c|c|}
		\hline
	$L\cdot W$ $\left[km^{2}\right]$ & $\Delta\sigma$ $\left[MPa\right]$ & $N_{tot}$   \\
	\hline
  4 & 510.7 & $overflow$ \\
  25 & 32.7 & 597.6 \\
  100 & 4.1 & 223.2\\
  225 & 1.2 & 107.8 \\
  \hline
	  \end{tabular}
	\end{center}
   \label{tab:Mfisso}
\end{table*}

\begin{table*}[!ht]
\caption{Results obtained for the best-fit of the sets of data reported in Figure 10 by the relation (\ref{eq:OmInt}). The symbol $rms$ indicates the standard deviation.}
	\begin{center}
		\begin{tabular}{c|cccc}
		  Slice &
		 $a$ $\left[y^{p-1}\right]$& $c$$\left[y\right]$& $p$& $rms$\\
		\hline
	$(0\div 0.5)km$ & $3.24\pm0.17$ &$0.37\cdot 10^{-7}\pm0.2\cdot 10^{-6}$& $1.02\pm0.02$ &1.86  \\
  $(0\div 8.5)km$ & $11.0\pm0.76$ &$0.23\cdot 10^{-7}\pm0.2\cdot 10^{-6}$& $0.91\pm0.032$ &$9.16$ \\
  $(0.5\div 8.5)km$ & $83.7\pm70.0$  &$3.1\pm1.96$ & $1.44\pm0.19$  &3.3  \\
  $(0\div 20)km$ & $34.3\pm8.0$  & $0.13\cdot10^{-1}\pm0.019$ &$1.14\pm0.077$  &9.22  \\
  $(0\div 100)km$ & $36.8\pm9.1$  &$0.15\cdot 10^{-1}\pm0.023$ & $1.12\pm0.08$  &10.8  \\
  $(20\div 100)km$ &$12.5\cdot 10^{3}\pm10.9\cdot 10^{3}$ & $61.0\pm5.43$ & $2.8\pm0.96\cdot 10^{-3}$ & $0.45$ \\
	  \end{tabular}
	\end{center}
  \label{tab:rgromori}
\end{table*}

\bibliographystyle{agu04}
\bibliography{Biblio}

\section*{Figure captions}

\indent

\textbf{Figure (1):} Conventions adopted for the reference system and for the parameters of the orientation of a fault: the strike $\phi$, dip $\delta$ and rake $\lambda$ angles and the unit vectors \textbf{n} and \textbf{l}.
The $x_{3}$ axis is positive towards the Nadir direction.\\ \vskip 1.0cm
\textbf{Figure (2):} Shear stress, $\sigma_{12}$, parallel to the slip direction for a strike-slip fault projected on the fault plane ($x_{1}x_{3}$, $x_{2}=0.2$ km).\\ \vskip 1.0cm
\textbf{Figure (3):} Tractions normal to the fault plane, $T_{n}$, projected on the plane $x_{1}x_{3}$ (the same plane of the fault; ($x_{2}=0.2$ km).\\ \vskip 1.0cm
\textbf{Figure (4):} Shear stress, $\sigma_{12}$, parallel to the slip direction for a strike-slip fault on a plane perpendicular to the fault ($x_{1}x_{2}$).\\ \vskip 1.0cm
\textbf{Figure (5):} Tractions normal to the fault plane, $T_{n}$, projected on the plane perpendicular to the fault ($x_{1}x_{2}$) .\\ \vskip 1.0cm
\textbf{Figure (6):} Coulomb stress change, $\Delta\tau$, projected on the plane perpendicular to the fault ($x_{1}x_{2}$). It is a combination of $\sigma_{12}$ and  $T_{n}$ on the same plane. \\ \vskip 1.0cm
\textbf{Figure (7):} Distribution of the triggered seismicity against time for different values of induced shear stress. The values of the parameters used are: $r=2$ $events/y$, $\Delta\tau=3$ $MPa$, $A\sigma=0.24$ and $\dot{\tau}=5$ $KPa/y$.\\ \vskip 1.0cm
\textbf{Figure (8):} Comparison between the behaviour of the rate against time in the two cases treated by  \citet{dieterich94} and the case that we have selected in our study.\\ \vskip 1.0cm
\textbf{Figure (9):} Distribution of the number of expected events in four different cases: (a) for a positive value of earthquake shear stress equal to 5 MPa, (b) for a negative value of it equal to -5 MPa, (c) the sum of (a) and (b) and (d) the integral of (c) over time. The values of $r$ and $t_{a}$ are fixed at $1$.\\ \vskip 1.0cm
\textbf{Figure (10):} Rappresentation of the total number of induced events (on vertical axis) in six different arbitrary slices (red, green and blue columns) around the principal fault after different intervals of time (on horizontal axis). The distances of the limits of the slices are computed from the perimeter of the source and the reference magnitude in the simulation is 6.0.
Histogram (a) rapresents the distribution nearest to the source at shorter intervals of time (the time scale is purely qualitative). Histogram (b) rapresents the distribution farther from the fault at longer intervals of time.\\ \vskip 1.0cm
\textbf{Figure (11):} Spatial distribution of the density of induced events at t=0.01 year.\\ \vskip 1.0cm
\textbf{Figure (12):} Spatial distribution of the density of induced events at t=1 year.\\ \vskip 1.0cm
\textbf{Figure (13):} Spatial distribution of the density of induced events at t=100 year.\\ \vskip 1.0cm
\textbf{Figure (14):} Comparison between the temporal distribution of the cumulative number of induced events
corresponding to synthetic data and the best fits obtained by the Omori law for the closer slices from the source.\\ \vskip 1.0cm
\textbf{Figure (15):} Comparison between the temporal distribution of the cumulative number of induced events
corresponding to synthetic data and the best fits obtained by the Omori law for the farther slices from the source.\\ \vskip 1.0cm

\end{document}